\documentclass[aip,graphicx]{revtex4-1}

\draft 
\usepackage{graphicx}
\begin{document}


\title{First-principles study on the magnetic properties of six potential half-metallic ferromagnets: alkaline-earth (Ca, Sr) doped $X$C ($X$$=$Si, Ge, Sn)} 



\author{Xiao-Ping Wei}
\author{Xian-Ru Hu}
\author{Hong Deng}
\author{Shi-Bin Chu}
\author{Jian-Bo Deng}
\altaffiliation{Corresponding author. Email: dengjb@lzu.edu.cn}
\affiliation{Department of Physics, LanZhou University, Lanzhou 730000, People's Republic of China}


\date{\today}

\begin{abstract}
Six half-metallic ferromagnets $X_{0.75}$$Y_{0.25}$C ($X$$=$Si, Ge, Sn and $Y$$=$Ca and Sr) with zinc-blende structure, resulting from alkaline-earth (Ca, Sr) substitution for $X$, are predicted based on the density functional theory. The calculated total magnetic moments of these ferromagnets are all integer 2.00$\mu_{B}$ per supercell, which are one of important characters of half-metallic ferromagnets. Our calculations indicate that $X_{0.75}$$Y_{0.25}$C have wide spin gap and potentially have high Curie temperature. Alkaline-earth doping results in the spin-polarization and half-metallicity of these compounds. It is confirmed that the $p$-$d$ exchange coupling is responsible for the ferromagnetism of $X_{0.75}$$Y_{0.25}$C except Sn$_{0.75}$Ca$_{0.25}$C.
\end{abstract}


\maketitle 

Spintronics, as potential materials for second-generation electronics, focused on the transmission of  charge and spin of electrons. It has more extensive application prospects than microelectronics as the first-generation electronics that only studies charges of electrons.\cite{S.A, Y, X} In recent years, there has been the increasing interest in the spintronic materials, especially the half-metallic ferromagnets (HMFs) which has a $\pm100\%$ spin-polarization at the Fermi level.\cite{S.K, X.P, J.C, J.W, S, J, V, X.F} HMFs possibly have higher magnetoresistance effect than ordinary magnetic materials due to their much higher spin-polarization at the Fermi level.\cite{H, B} Therefore, HMFs will have much wider application as spintronics materials. Many high-quality spintronic devices based on the magnetoresistance effect of HMFs, such as magnetoresistance random access memoery and read-write magnetic head of computers, may be invented in the near future. The realization of semiconductor spintronic devices, however, requires efficient electrical spin injection from magnetic electrodes to semiconductors.\cite{J.C, J.W} HMFs have 100\% spin polarization, and their resistivity matches well with that of semiconductors so that they are considered as the most suitable magnetic electrode materials of spin semiconductors. For above mentioned reasons, HMFs have attracted much more attention and become one of the hottest topics in the past years. Especially, dilute magnetic semiconductors (DMSs), such as Mn (Cu)-doped GaN and Cr-doped ZnSe, AlN, ZnTe etc.,\cite{L, R.Q, H.X, H.S} which has a few fractions of host elements usually replaced by magnetic ions. It is reported that the doping using magnetic elements is confronted with these problems related to precipitates or secondary phase formation, which is undesirable in practical applications.\cite{J.H, T, S.Q} Therefore, it may be meaningful to consider unconventional doping elements except for magnetic elements. Recently, some groups reported the intrinsic nonmagnetic elements, such as C, N, Mg and Ca, are treated as dopants favoring spin polarization and ferromagnetic coupling in semiconductors.\cite{S.P, C, O} However, few works have been focused on alkaline-earth elements (Ca, Sr) doped $X$C ($X$$=$Si, Ge, Sn) but Si$_{0.75}$Ca$_{0.25}$C.\cite{H.M} In this paper, six potential HMFs, alkaline-earth doped $X$C with the zinc-blende structure, are predicted. Moreover their electronic and magnetic properties are studied based on the density functional theory.\par
Generally, the experimental dopant concentration is from 5\% to 30\% for compounds with zinc-blende structures. In order to achieve the realistic dopant concentration, we used a periodic supercell $X$C ($X$$=$Si, Ge, Sn), which consists of four $X$ atoms and four C atoms. Substitution of one $X$ atom by one alkaline-earth atom, resulting in each alkaline-earth atom connects with four adjacent C atoms and three second nearest neighboring $X$ atoms in the periodic supercell, then the alkaline-earth dopant concentration is 25\%. The supercell of alkaline-earth $Y$ (Ca, Sr) doped $X$C is shown in Fig. 1.\par
 The calculations of magnetic and electronic properties are performed using
 the scalar relativistic version of the full-potential local-orbital (FPLO) minimum-basis method.\cite{K, I} In this scheme, the scalar relativistic Dirac equation was solved. For the present calculations, the site-centered potentials and densities were expanded in spherical harmonic contributions up to $\ell_{max}$ =12. The Perdew-Burke-Ernzerhof 96 of the generalized gradient approximation (GGA) was used for exchange-correlation (XC) potential. \cite{J.P}
Accurate Brillouin zone integrations were performed using the standard special k point technique of the tetrahedron. We found that 30$\times$30$\times$30$=$27000 k points were sufficient in all cases. For a self-consistent field iteration, the charge density is converged to 10$^{-6}$, which corresponds to a total energies (10$^{-8}$ hartree).\par

Some magnetic and electronic parameters of $X_{0.75}$$Y_{0.25}$C, including the lattice constants (a${_0}$$=$b${_0}$$=$c${_0}$), local magnetic moments, spin gap ($E{_b}$), half-metallic gap ($E{_g}$) and the supercell net magnetic moments ($M$), are calculated and shown in Table 1. The lattice constants of $X_{0.75}$$Y_{0.25}$C are larger than those of $X$C. The main reason is that covalent radius of alkaline-earth ions larger than that of $X$ ions. Therefore, the supercell will expand when $X$ ions are substituted by one alkaline-earth ions. However, the lattice constants of $X_{0.75}$Ca$_{0.25}$C are smaller than those of $X_{0.75}$Sr$_{0.25}$C. Similarly, much larger ion radius of Sr-ion compared to Ca-ion also results in expansion of the supercell, but the supercell will be compressed more seriously due to Ca-ion much smaller ion radius when the $X$-ions is substituted by one Ca-ion.\par
The spin-polarized density of states of $X_{0.75}$$Y_{0.25}$C ($X$$=$Si, Ge, Sn and $Y$$=$Ca, Sr) are shown in Fig. 2. There is obvious spin polarization for $X_{0.75}$$Y_{0.25}$C near the Fermi level while there is no spin-polarization for $X$C. This means that alkaline-earth doping can exert substantial influence on the electronic and magnetic properties of $X_{0.75}$$Y_{0.25}$C. It is very important that there are spin gap only for spin-up subbands at the Fermi level, so these six materials are all potential HMFs. Furthermore, the net magnetic moments of supercell are the integer 2.00$\mu{_B}$, which is one of half-metallic characters of these materials.\par
From Table 1 and Fig. 2, the spin gaps defined as the energy distance from the valence band maximum to the conduction band minimum for the spin-up subband. It can be seen that the spin gaps of these HMFs are wide and they possibly have high Curie temperature, which is probably close to or higher than many zinc blende or wurtzite HMFs.\cite{W, M, L.K} This is very important for application of magnetic materials. On the other hand, the half-metallic gaps or the spin-flip gaps, defined as the energy difference between the Fermi level of the metallic highest occupied subbands and the peak of the semiconducting lowest unoccupied subbands. Generally, the values of spin-flip gaps are underestimated compared to experimental half-metallic gaps. Therefore, we think the half-metallicity of these HMFs is stable, among of them Si$_{0.75}$Sr$_{0.25}$C is most stable.\par
From the crystal field theory, isolated alkaline-earth (Ca, Sr) have weaker electronegativity than isolated C atoms or $X$ atoms, so there are both covalent bonds and ionic bonds between the alkaline-earth ions and C-ions in $X_{0.75}$$Y_{0.25}$C, but only covalent bonds in $X$C. It is evident that there are the presence of $p$-$d$ orbitals hybridization and $sp$ orbitals hybridization due to their close energy ($s$ states are not shown due to their little contribution), the gap basically arises from the covalent hybridization which forms the bonding and antibonding bands. The $p$-$d$ hybridization produces larger magnetic moments on the non-magnetic C sites, the total magnetic moment of $X_{0.75}$$Y_{0.25}$C mainly comprises the C component with small contributions from alkaline-earth (Ca, Sr) and $X$ ($X$$=$Si, Ge, Sn) atoms. We also note that the negative value of the magnetic moment for $X$ ($X$$=$Si, Ge, Sn), indicating antiferromagnetic interaction with other atoms. \par

Generally, the electronic structures calculations can be used to work out two important parameters, namely the $s$-$d$ exchange constants $N{_0}$$\alpha$ (conduction band) and the $p$-$d$ exchange constant $N{_0}$$\beta$ (valence band), where $N{_0}$ denotes the concentration of cations, these exchange constants can be determined directly from the valence and conduction band-edge spin splitting. The fundamental idea originates from mean field theory which is based on the Hamiltonian $H$$=$$-$\emph{N}${_0}$$\beta$$\emph{\textbf{s}}$$\cdot$$\textbf{\emph{S}}$,\cite{Sa, La} where $\beta$ is the $p$-$d$ exchange integral, while $s$ and $S$ are the hole spin and alkaline-earth spin, respectively. The exchange constants can be defined as

$N{_0}$$\alpha$$=$$\frac{\Delta E{_c}}{x <S>}$,
$N{_0}$$\beta$$=$$\frac{\Delta E{_v}}{x <S>}$

where $\Delta$$E$${_c}$ and $\Delta$$E$${_v}$ are the respective band edge splitting
of the conduction band maxima and valence band
maxima at the $\Gamma$ point, $x$ is the concentration of alkaline-earth in $X$C and $<$S$>$ is half of total magnetic moment.
 \cite{S.N, H.M} The calculated values of  $\Delta$$E$${_c}$, $\Delta$$E$${_v}$, $N{_0}$$\alpha$ and $N{_0}$$\beta$ are listed in Table II.
From our results except for Sn$_{0.75}$Ca$_{0.25}$C, it is confirmed that the $p$-type semiconductor is induced and the ferromagnetism is attributed to the $p$-$d$ exchange coupling. The mechanism of the stabilization of
ferromagnetism depends on the position of the impurity $d$ levels with respect to the valence band edge.\cite{G.M}\par

In summary, six HMFs with zinc-blende structure, $X_{0.75}$$Y_{0.25}$C ($X$$=$Si, Ge, Sn and $Y$$=$Ca, Sr), are predicted based on the density functional theory. They have high Curie temperature possibly and their magnetic moments are 2.00$\mu{_B}$. The main reason of spin polarization and half-metallicity of $X_{0.75}$$Y_{0.25}$C is that the covalent hybridization in $X_{0.75}$$Y_{0.25}$C leads to the formation
of bonding and antibonding bands with a gap. Except for Sn$_{0.75}$Ca$_{0.25}$C, the $p$-$d$ exchange coupling plays a crucial role in forming the ferromagnetism alkaline-earth doped $X$C.

\begin{acknowledgments}
The authors would thank to be grateful for W. X. Zhang, H. M. Huang, X. M. Tao and Ulrike Nitzsche for many useful discussions.\par
\end{acknowledgments}

\begin{figure}
 \centering
  \includegraphics[width=12cm]{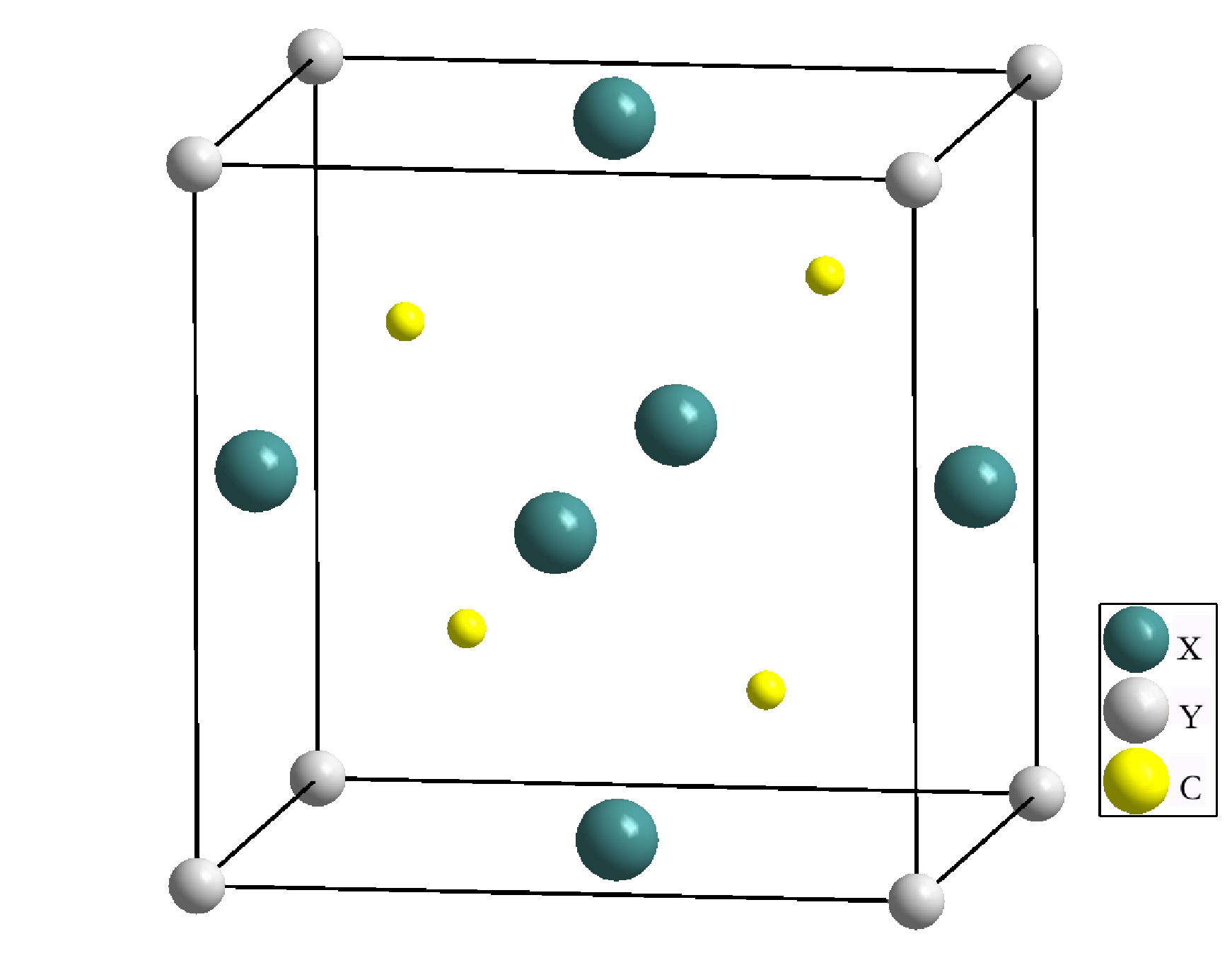}\\
  \caption{The supercell structure of $X_{0.75}$$Y_{0.25}$C}\label{figure}
\end{figure}

\begin{figure}
  \centering
  \includegraphics[width=13cm,angle=-90]{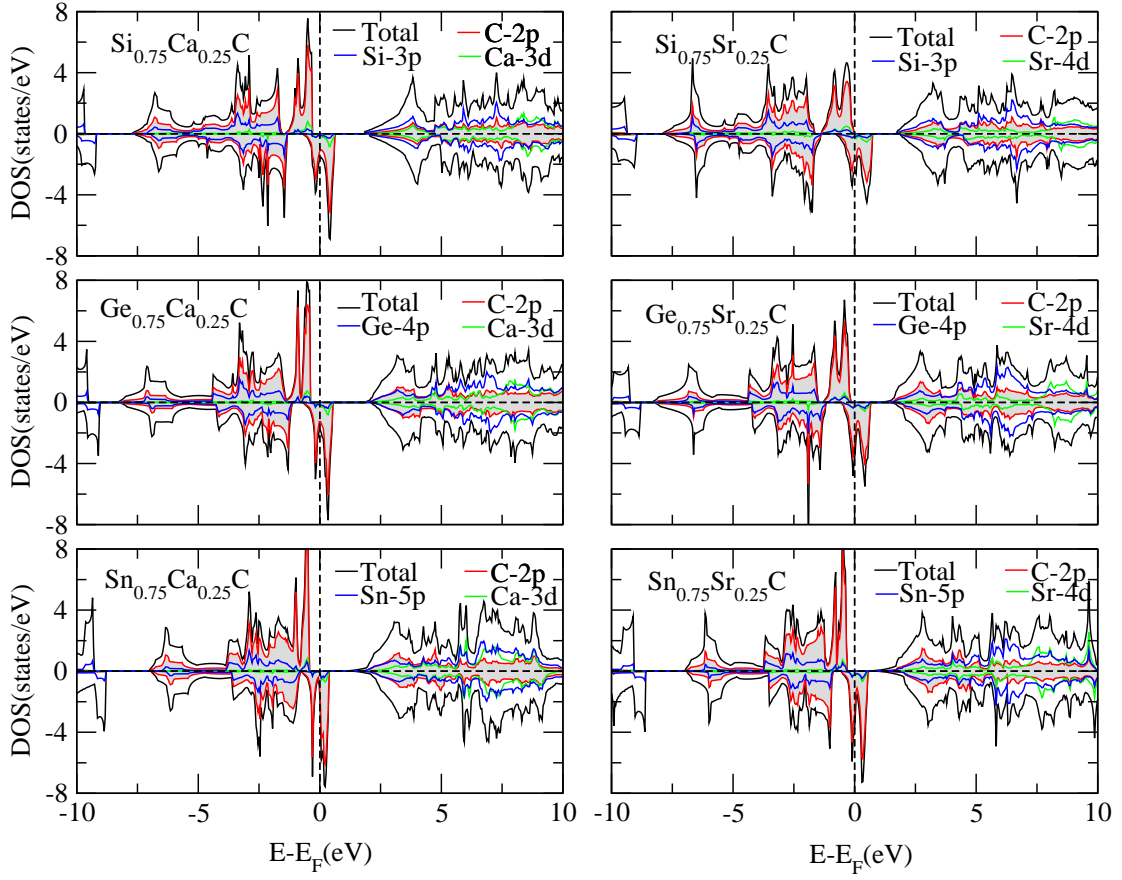}\\
  \caption{Spin-polarized density of states of six potential HMFs $X_{0.75}$$Y_{0.25}$C ($X$$=$Si, Ge, Sn and $Y$$=$Ca, Sr). Positive value represents spin-up density of states, otherwise spin-down density of states.}\label{figure}
\end{figure}

\begin{table}[b]
\caption{\label{tab:test}
Some calculated parameters of $X_{0.75}$$Y_{0.25}$C ($X$$=$Si, Ge, Sn and $Y$$=$Ca, Sr) are listed.}
\begin{ruledtabular}
\begin{tabular}{cccccccc}
 Supercell & $a{_0}$ & $M_{X}$ & $M_{Y}$ & $M_{C}$ & $M$& $E_{b}$ &$E_{g}$ \\
           & (${\AA}$) & ($\mu{_B}$) & ($\mu{_B}$) & ($\mu{_B}$) &($\mu{_B}$)& (eV) & (eV)\\
\hline
  Si${_3}$CaC${_4}$ & 4.68 & -0.036  & 0.118 & 0.497 &2.00 & 2.14 & 2.13  \\
  Si${_3}$SrC${_4}$ & 4.80 & -0.025  & 0.140 & 0.484 &2.00 & 1.71 & 2.47  \\
  Ge${_3}$CaC${_4}$ & 4.88 & -0.056  & 0.101 & 0.516 &2.00 & 2.15 & 1.86  \\
  Ge${_3}$SrC${_4}$ & 5.01 & -0.047  & 0.095 & 0.512 &2.00 & 1.08 & 1.36  \\
  Sn${_3}$CaC${_4}$ & 5.24 & -0.064  & 0.067 & 0.531 &2.00 & 1.41 & 2.19 \\
  Sn${_3}$SrC${_4}$ & 5.34 & -0.057  & 0.055 & 0.529 &2.00 & 0.83 & 0.68  \\
\end{tabular}
\end{ruledtabular}

\end{table}

\begin{table}[b]
\caption{\label{tab:test}
Calculated band-edge spin splitting for conduction band $\Delta$$E{_c}$ and valence band $\Delta$$E{_v}$ and $N{_0}$$\alpha$ and $N{_0}$$\beta$ are the exchange constants  for $X_{0.75}$$Y_{0.25}$C ($X$$=$Si, Ge, Sn and $Y$$=$Ca, Sr) compounds.}
\begin{ruledtabular}
\begin{tabular}{ccccc}
 Supercell & $\Delta$$E{_c}$ (eV)& $\Delta$$E{_v}$ (eV)& $N{_0}$$\alpha$ (eV)& $N{_0}$$\beta$ (eV)\\
 \hline
  Si${_3}$CaC${_4}$ & 0.215 & 0.627  & 0.862 & 2.509  \\
  Si${_3}$SrC${_4}$ & 0.235 & 0.608  & 0.940 & 2.431   \\
  Ge${_3}$CaC${_4}$ & 0.394 & 0.583  & 1.576 & 2.333   \\
  Ge${_3}$SrC${_4}$ & 0.410 & 0.559  & 1.639 & 2.237   \\
  Sn${_3}$CaC${_4}$ &-0.910 &-0.753  &-3.640 &-3.013  \\
  Sn${_3}$SrC${_4}$ & 0.389 & 0.503  & 1.557 & 2.014   \\
\end{tabular}
\end{ruledtabular}

\end{table}

\end{document}